\documentclass[sigconf,table,xcdraw,screen]{acmart}
\usepackage[ruled,vlined]{algorithm2e}
\usepackage{graphicx}
\usepackage{caption}
\usepackage{subcaption}
\usepackage{paralist}
\usepackage{tabularx}
\usepackage{colortbl}
\usepackage{balance}
\usepackage{multirow}
\usepackage{multicol}
\usepackage{pbox}
\usepackage{mathtools}
\usepackage{algorithmic}
\usepackage{paralist}
\usepackage{tabularx}
\usepackage{url}
\usepackage{balance}
\usepackage{multirow}
\usepackage{hyperref} 
\usepackage{multicol}
\usepackage{amsmath}
\usepackage{setspace}
\usepackage{verbatim}
\usepackage{float}
\usepackage[normalem]{ulem}
\usepackage{xspace}
\usepackage{amssymb}
\usepackage{ifthen}
\usepackage{pifont}
\usepackage{tikz}
\usepackage{hyperref}
\usepackage[most]{tcolorbox}

\definecolor{MidnightBlue}{HTML}{006895}
\newcommand{\vts}{{\sc V2S}\xspace}
\newcommand{\vtss}{{\sc V2S's~}}

\newcommand{\ReDraw}{{\sc ReDraw}\xspace}

\newcommand{\approach}{{\sc V2S}\xspace}
\newcommand{\approachs}{{\sc V2S's~}}
\newcommand{\Faster}{{\sc Faster R-CNN}\xspace}

\newcommand{\Rcnn}{{\sc R-CNN}\xspace}
\newcommand{\Cnn}{{\sc CNN}\xspace}
\newcommand{\Cnns}{{\sc CNNs}\xspace}
\newcommand{\AlexNet}{{\sc AlexNet}\xspace}

\newcommand{\VGGNet}{{\sc VGGNet}\xspace}

\include{macro}

\AtBeginDocument{%
  \providecommand\BibTeX{{%
    \normalfont B\kern-0.5em{\scshape i\kern-0.25em b}\kern-0.8em\TeX}}}

\copyrightyear{2020}
\acmYear{2020}
\setcopyright{acmcopyright}
\acmConference[ICSE '20]{42nd International Conference on Software Engineering}{May 23--29, 2020}{Seoul, Republic of Korea}
\acmBooktitle{42nd International Conference on Software Engineering (ICSE '20), May 23--29, 2020, Seoul, Republic of Korea}
\acmPrice{15.00}
\acmDOI{10.1145/3377811.3380328}
\acmISBN{978-1-4503-7121-6/20/05}

\begin{document}

\title{Translating Video Recordings of \\Mobile App Usages into Replayable
Scenarios\\}

\author{Carlos Bernal-C\'{a}rdenas}
\affiliation{
  \institution{William \& Mary}
  \city{Williamsburg}
  \state{Virginia}
  \country{USA}
}
\email{cebernal@cs.wm.edu}

\author{Nathan Cooper}
\affiliation{
  \institution{William \& Mary}
  \city{Williamsburg}
  \state{Virginia}
  \country{USA}
}
\email{nacooper01@email.wm.edu}

\author{Kevin Moran}
\affiliation{
  \institution{William \& Mary}
  \city{Williamsburg}
  \state{Virginia}
  \country{USA}
}
\email{kpmoran@cs.wm.edu}

\author{Oscar Chaparro}
\affiliation{
  \institution{William \& Mary}
  \city{Williamsburg}
  \state{Virginia}
  \country{USA}
}
\email{oscarch@wm.edu}

\author{Andrian Marcus}
\affiliation{
  \institution{The University of Texas at Dallas}
  \city{Dallas}
  \state{Texas}
  \country{USA}
}
\email{amarcus@utdallas.edu}

\author{Denys Poshyvanyk}
\affiliation{
  \institution{William \& Mary}
  \city{Williamsburg}
  \state{Virginia}
  \country{USA}
}
\email{denys@cs.wm.edu}

\renewcommand{\shortauthors}{C. Bernal-C\'{a}rdenas, N. Cooper, K. Moran, O. Chapparo, A. Marcus \& D. Poshyvanyk}
\renewcommand{\shorttitle}{Translating Video Recordings of Mobile App Usages into Replayable Scenarios}

\begin{abstract}

Screen recordings of mobile applications are easy to obtain and capture a wealth of information pertinent to software developers (\eg bugs or feature requests), making them a popular mechanism for crowdsourced app feedback. Thus, these videos are becoming a common artifact that developers must manage. In light of unique mobile development constraints, including swift release cycles and rapidly evolving platforms, automated techniques for analyzing all types of rich software artifacts provide benefit to mobile developers. Unfortunately, automatically analyzing screen recordings presents serious challenges, due to their graphical nature, compared to other types of (textual) artifacts. To address these challenges, this paper introduces \vts, a lightweight, automated approach for translating video recordings of Android app usages into replayable scenarios. \vts is based primarily on computer vision techniques and adapts recent solutions for object detection and image classification to detect and classify user actions captured in a video, and convert these into a replayable test scenario. We performed an extensive evaluation of \vts involving 175 videos depicting 3,534 GUI-based actions collected from users exercising features and reproducing bugs from over 80 popular Android apps. Our results illustrate that \vts can accurately replay scenarios from screen recordings, and is capable of reproducing $\approx$ 89\% of our collected videos with minimal overhead. A case study with three industrial partners illustrates the potential usefulness of \vts from the viewpoint of developers. 
\vspace{-0.2cm}
\end{abstract}



\maketitle

\vspace{-0.3cm}
\section{Introduction}
\label{sec:intro}

Mobile application developers rely on a diverse set of software artifacts to help them make informed decisions throughout the development process. These information sources include user reviews, crash reports, bug reports, and emails, among others. An increasingly common component of these software artifacts is graphical information, such as screenshots or screen recordings. This is primarily due to the fact that they are relatively easy to collect and, due to the GUI-driven nature of mobile apps, they contain rich information that can easily demonstrate complex concepts, such as a bug or a feature request. 
In fact, many crowd-testing and bug reporting frameworks have built-in screen recording features to help developers collect mobile application usage data and faults~\cite{watchsend,bugclipper,appsee,testfairy}. 
\textit{Screen recordings} that depict application usages are used by developers to: (i) help understand how users interact with apps~\cite{Schusteritsch:CHI'07,mrtappy}; (ii) process bug reports and feature requests from end-users~\cite{3Bettenburg:FSE08}; and (iii) aid in bug comprehension for testing related tasks~\cite{Mao:ASE17}. 
However, despite the growing prevalence of visual mobile development artifacts, developers must still manually inspect and interpret screenshots and videos in order to glean relevant information, which can be time consuming and ambiguous. 
The manual effort required by this comprehension process complicates a development workflow that is already constrained by language dichotomies~\cite{Moran:ICPC'18} and several challenges unique to mobile software, including: (i) pressure for frequent releases~\cite{Hu:ESYS14,Jones:2014}, (ii) rapidly evolving platforms and APIs~\cite{Linares:FSE13,Bavota:TSE15}, (iii) constant noisy feedback from users~\cite{Ciurumelea:SANER'17,DiSorbo:FSE'16,Palomba:ICSE17,Palomba:JSS'18,Palomba:ICSME15}, and (iv) fragmentation in the mobile device ecosystem~\cite{Han:WCRE'12,Wei:ASE'16,android-fragmentation} among others~\cite{Linares-Vasquez:ICSME'17B}. Automation for processing graphical software artifacts is necessary and would help developers shift their focus toward core development tasks.

To improve and automate the analysis of video-related mobile development artifacts, we introduce \NEW{Video to Scenario (\vts)}, a lightweight automated approach for translating video screen recordings of Android app usages into replayable scenarios. We designed \vts to operate solely on a video file recorded from an Android device, and as such, it is based primarily on computer vision techniques. \vts adapts recent Deep Learning (DL) models for object detection and image classification to accurately detect and classify different types of user actions performed on the screen. These classified actions are then translated into replayable scenarios that can automatically reproduce user interactions on a target device, making \vts the first purely graphical Android record-and-replay technique.

In addition to helping automatically process the graphical data that is already present in mobile development artifacts, \vts can also be used for improving or enhancing additional development tasks that do not currently take full advantage of screen-recordings, such as: creating and maintaining automated GUI-based test suites; and crowdsourcing functional and usability testing. 

We conducted a comprehensive evaluation of \vts using both videos collected from users reproducing bugs as well as general usage videos from the top-rated apps of 32 categories in the Google Play market. As part of this evaluation, we examined the effectiveness of the different components that comprise \vts as well as the accuracy of the generated scenarios. Additionally, we assessed the overhead of our technique and  
conducted a case study with three industrial partners to understand the practical applicability of \vts. The results of our evaluation indicate that \vts is \textit{accurate}, and is able to correctly reproduce 89\% of events across collected videos. The approach is also \textit{robust} in that it is applicable to a wide range of popular native and non-native apps currently available on Google Play. In terms of \textit{efficiency}, we found that \vts imposes acceptable overhead, and is perceived as potentially useful by developers.

In summary, the main contributions of our work are as follows:

\begin{itemize}
	\item{\vts, the first record-and-replay approach for Android that functions purely on screen-recordings of app usages. \vts adapts computer vision solutions for object detection, and image classification, to effectively recognize and classify user actions in the video frames of a screen recording;}
	\item{\NEW{An automated pipeline for dataset generation and model training to identify user interactions from screen recordings;}}
	\item{The results of an extensive empirical evaluation of \vts that measures the \textit{accuracy}, \textit{robustness}, and \textit{efficiency} across 175 videos from 80 applications;} 
	\item{The results of a case study with three industrial partners, who develop commercial apps, highlighting \vtss potential usefulness, as well as areas for improvement and extension;} 	\item{An online appendix~\cite{appendix}, which contains examples of videos replayed by \vts, experimental data, source code, trained models, and our evaluation infrastructure to facilitate reproducibility of the approach and the results.}
\end{itemize}

\section{Background}
\label{sec:background}

We briefly discuss DL techniques for image classification and object detection that we adapt for touch/gesture recognition in \vts. 

\vspace{-0.2cm}
\subsection{Image Classification}
\label{subsec:back-CNNs}

Recently, DL techniques that make use of neural networks consisting of specialized layers have shown great promise in classifying diverse sets of images into specified categories. Advanced approaches leveraging Convolutional Neural Networks (CNNs) for highly precise image recognition~\cite{Krizhevsky:NIPS12, Simonyan:ICLR14,Zeiler:ECCV14,Szegedy:CVPR15,He:CVPR16} have reached human levels of accuracy for image classification tasks. 

Typically, each \textit{layer} in a \Cnn performs some form of computational transformation to the data fed into the model. The initial layer usually receives an input image. This layer is typically followed by a convolutional layer that extracts features from the pixels of the image, by applying \textit{filters} (\aka kernels) of a predefined size, wherein the contents of the filter are transformed via a pair-wise matrix multiplication (\ie the convolution operation). Each filter is passed throughout the entire image using a fixed \textit{stride} as sliding window to extract \textit{feature maps}. Convolutional layers are used in conjunction with ``max pooling'' layers to further reduce the dimensionality of the data passing through the network. The convolution operation is linear in nature. Since images are generally non-linear data sources, activation functions such as Rectified Linear Units (ReLUs) are typically used to introduce a degree of non-linearity. Finally, a fully-connected layer (or series of these layers) are used in conjunction with a \textit{Softmax classifier} to predict an image class. The training process for CNNs is usually done by updating the weights that connect the layers of the network using gradient descent and back-propagating error gradients. 

\vts implements a customized CNN for the specific task of classifying the opacity of an image to help segment GUI-interactions represented by a touch indicator (see Section \ref{subsec:approach-detection}).

\subsection{Object Detection}
\label{subsec:back-RCNNs}

\begin{figure}[t]
    \begin{center}
		\includegraphics[width=\columnwidth]{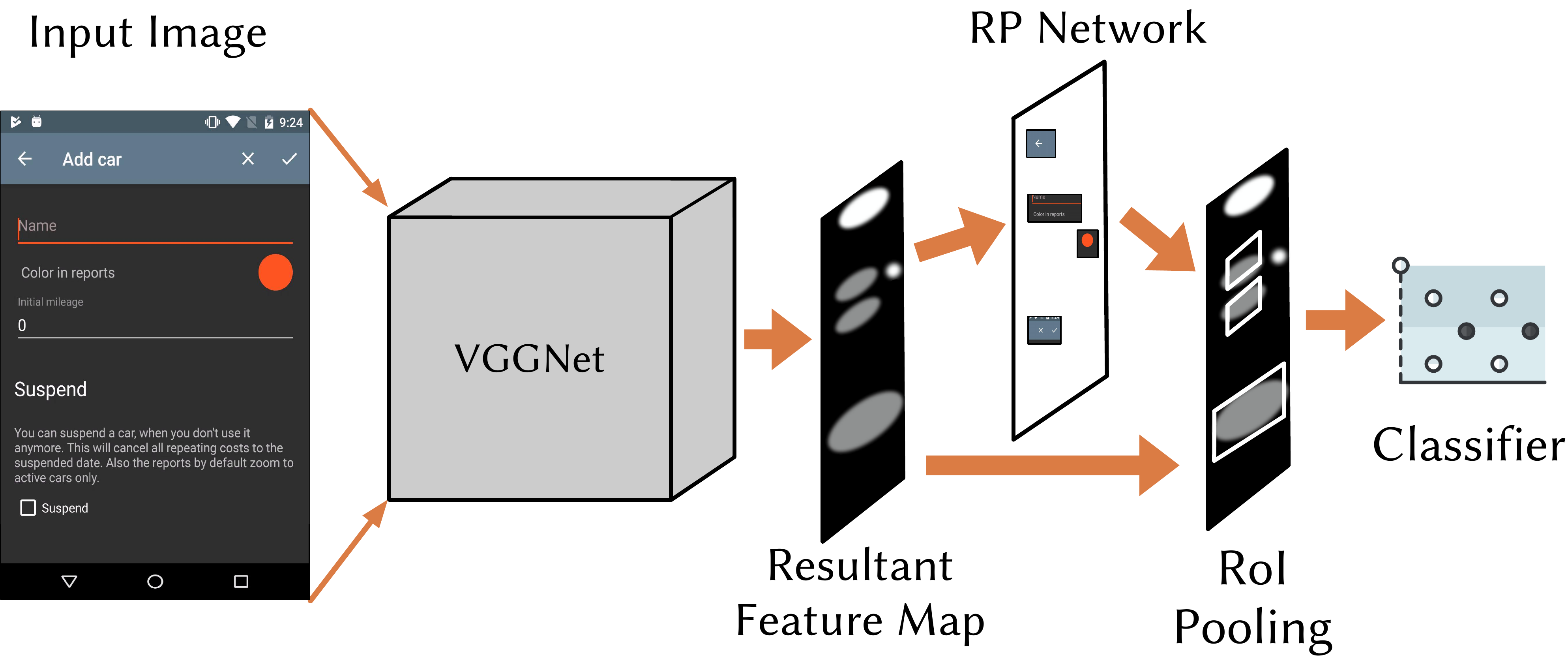}
		\vspace{-0.4cm}
        \caption{Illustration of the \Faster Architecture}
        \label{fig:faster-rcnn}
        \vspace{-0.8cm}
    \end{center}
\end{figure}

In the task of image classification, a single, usually more general label (\eg \textit{bird} or \textit{person}) is assigned to an entire image. However, images are typically multi-compositional, containing different objects to be identified. 
Similar to image classification, DL models for object detection have advanced dramatically in recent years, enabling object tracking and counting, as well as face and pose detection among other applications. One of the most influential neural architectures that has enabled such advancements is the \Rcnn introduced by Girshick \etal~\cite{Girshick:CVPR14}. The \Rcnn architecture combines algorithms for image region proposals (RPs), which aim to identify image regions where content of interest is likely to reside, with the classification prowess of a \Cnn. An \Rcnn generates a set of RP bounding-boxes using a selective search algorithm~\cite{Uijlings:IJCV'13}. Then, all identified image regions are fed through a pre-trained \AlexNet~\cite{Krizhevsky:NIPS12} (\ie the \textit{extractor}) to extract image features into vectors. These vectors are fed into a support vector machine (\ie the \textit{classifier}) that determines whether or not each image region contains a class of interest. Finally, a greedy non-maximum suppression algorithm (\ie the \textit{regressor}) is used to select the highest likelihood, non-overlapping regions, as classified objects.

\begin{figure*}[t]
    \begin{center}
	
		\includegraphics[width=0.93\linewidth]{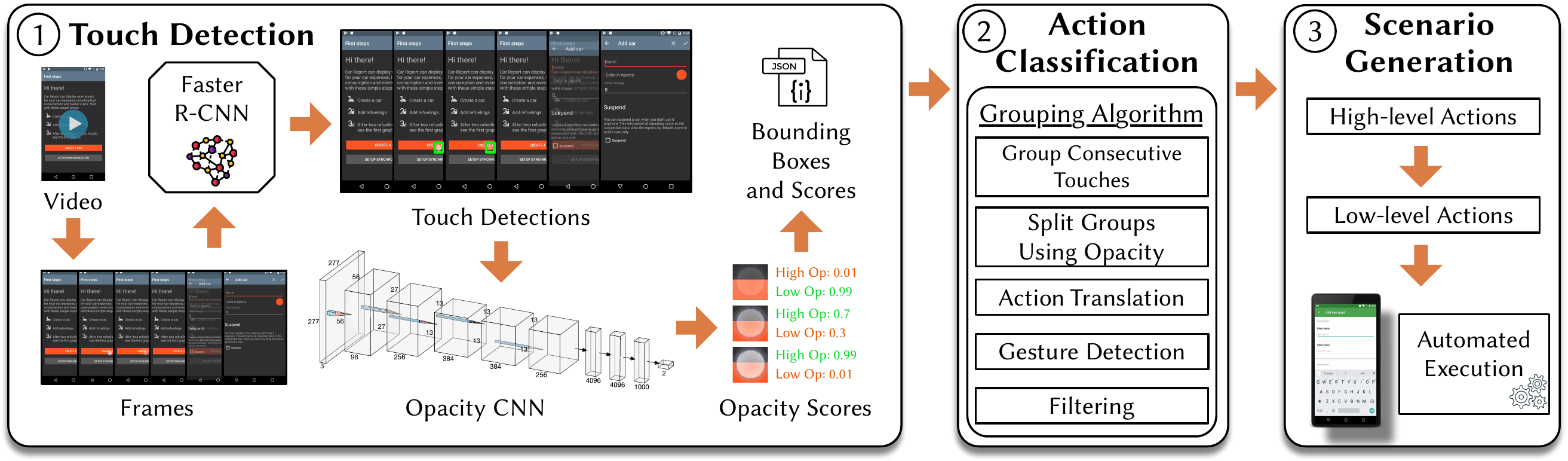}
		\vspace{-0.3cm}
        \caption{The \vts Approach and Components}
        \label{fig:approach}
        \vspace{-0.2cm}
    \end{center}
\end{figure*}

In this paper, we utilize the the \Faster~\cite{Ren:NIPS15} architecture~(Fig. \ref{fig:faster-rcnn}), which improves upon \Rcnn architecture through the introduction of a separate NN to predict image region proposals. By integrating the training of the region proposal network into the end-to-end training of the network, both the speed and accuracy of the model increase. In \vts, we adapt the \Faster model to detect a touch indicator representing a user action in video frames.

\section{The V2S Approach}
\label{sec:approach}

This section outlines the \vts approach for automatically translating Android screen recordings into replayable scenarios. \figref{fig:approach} depicts \vtss architecture, which is divided into three main phases: (i) the \textit{Touch Detection} phase, which identifies user touches in each frame of an input video; (ii) the \textit{Action Classification} phase that groups and classifies the detected touches into discrete user actions (\ie tap, long-tap, and swipe), and (iii) the \textit{Scenario Generation} phase that exports and formats these actions into a replayable script. Before discussing each phase in detail, we discuss some preliminary aspects of our approach, input specifications, and requirements.

\subsection{Input Video Specifications}
\label{subsec:approach-prelim}

\begin{figure}[t]
    \begin{center}
		\includegraphics[width=0.6\columnwidth]{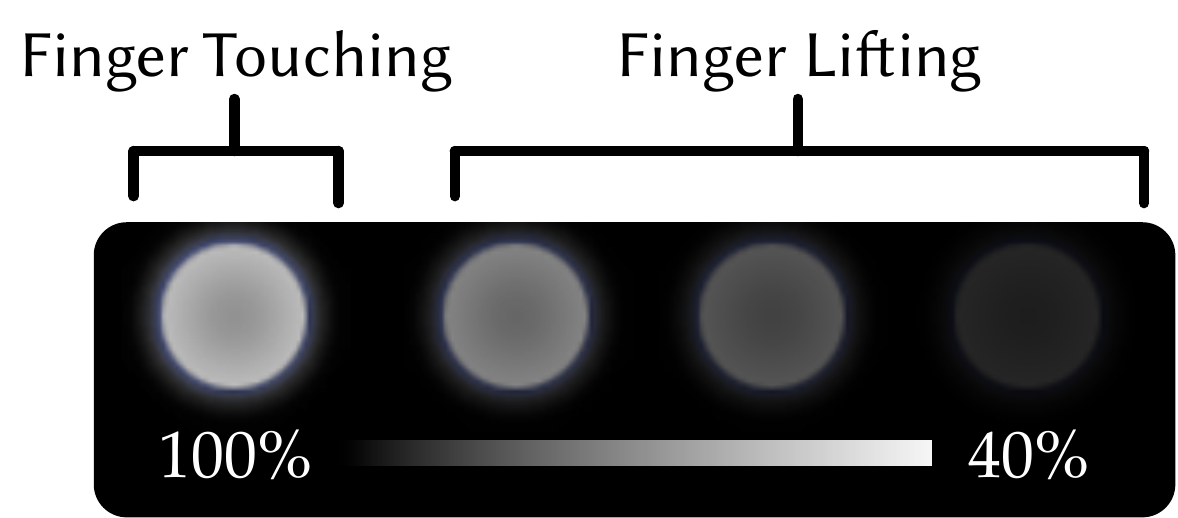}
		\vspace{-0.2cm}
        \caption{Illustration of touch indicator opacity levels}
        \label{fig:touch-opacity}
        \vspace{-0.7cm}
    \end{center}
\end{figure}

	In order for a video to be consumable by \vts, it must meet a few lightweight requirements to ensure proper functioning with our computer vision (CV) models. First, the video frame size must match the full-resolution screen size of the target Android device, in order to be compatible with a specified pre-trained object-detection network. This requirement is met by nearly every modern Android device that has shipped within the last few years. These videos can be recorded either by the built-in Android \texttt{screenrecord} utility, or via third-party applications~\cite{g-play-recording-apps}. The second requirement is that input videos must be recorded at least 30 ``frames per second'' (FPS), which again, is met or exceeded by a majority of modern Android devices. This requirement is due to the fact that the frame-rate directly corresponds to the accuracy with which ``quick'' gestures (\eg fast scrolling) can be physically resolved in constituent video frames. Finally, the videos must be recorded with the ``Show Touches'' option enabled on the device, which is accessed through an advanced settings menu~\cite{show-touches}, and is available by default on nearly all Android devices since at least Android 4.1.  
	This option renders a \textit{touch indicator}, which is a small semi-transparent circle, that gives a visual feedback when the user presses her finger on the device screen. The opacity of the indicator is fully solid from the moment the user first touches the screen and then fades from more to less opaque when a finger is lifted off the screen (\figref{fig:touch-opacity}).

\subsection{Phase 1: Touch Detection}
\label{subsec:approach-detection}

The \textit{goal} of this phase is to accurately identify the locations where a user touched the device screen during a video recording.  To accomplish this, \vts leverages the DL techniques outlined in Sec.~\ref{sec:background} to both accurately find the position of the touch indicator appearing in video frames, and identify its opacity to determine whether a user's finger is being pressed or lifted from the screen. More specifically, we adapt an implementation of \Faster~\cite{Ren:NIPS15,Ren:PAMI17}, which makes use of \VGGNet~\cite{Simonyan:ICLR14} for feature extraction of RPs in order to perform touch indicator detection. To differentiate between low and high-opacity detected touch indicators, we build an {\sc Opacity CNN}, which is a modified version of \AlexNet~\cite{Krizhevsky:NIPS12}. Given that we adapt well-known DL architectures, here we focus on describing our adaptions, and provide model specs in our appendix~\cite{appendix}.

The \textit{Touch Detection} Phase begins by accepting as input a video that meets the specifications outlined in \secref{subsec:approach-prelim}. First, the video is \textit{parsed} and decomposed into its constituent frames. Then the \Faster network is utilized to \textit{detect} the presence of the touch indicator, if any, in every frame. Finally, the {\sc Opacity CNN} \textit{classifies} each detected touch indicator as having either low or high-opacity. The output of this phase is a structured \texttt{JSON} with a set of touch indicator bounding boxes in individual video frames wherein each detected touch indicator is classified based on the opacity.

\subsubsection{Parsing Videos}
\label{subsubsec:appraoch-parsing}

Before \vts parses the video to extract single frames, it must first normalize the frame-rate for those videos where it may be variable, to ensure a constant FPS. Certain Android devices may record variable frame-rate video for efficiency~\cite{stackoverflow-android-screen-record}. This may lead to inconsistencies in the time between frames, which \vts utilizes in the classification phase to synthesize the timing of touch actions. Thus, to avoid this issue, we normalize the frame rate to 30fps and extract individual frames using the \texttt{FFmpeg}~\cite{ffmpeg} tool. 

\subsubsection{Faster R-CNN}
\label{subsubsec:approach-faster-rcnn}

After the individual frames have been parsed from the input video, \vts applies its object detection network to localize the bounding boxes of touch indicators. However, before using the object detection, it must be trained. As described in \secref{sec:background} the DL models that we utilize typically require large, manually labeled datasets to be effective. However, to avoid the manual curation of data, and  make the \vts approach practical, we designed a fully automated dataset generation and training process. To bootstrap the generation of \vtss object detection training dataset, we make use of the existing large-scale \ReDraw dataset of Android screenshots~\cite{Moran:TSE18}. This dataset includes over 14k screens extracted from the most popular Android applications on Google Play using a fully-automated execution technique.

Next, we randomly sample 5k unique screenshots of different apps and programmatically superimpose an image of the \textit{touch indicator} at a random location in each screenshot. During this process, we took two steps to ensure that our synthesized dataset reflects actual usage of the touch indicator: (i) we varied the opacity of the indicator icon between 40\%-100\% to ensure our model is trained to detect instances where a finger is lifted off the screen; (ii) we placed indicator icons on the edges of the screen to capture instances where the indicator may be occluded. This process is repeated three times per screenshot to generate 15k unique images. \NEW{We then split this dataset 70\%/30\% to create training and testing sets respectively. We performed this partitioning such that all screenshots expect one appear only in the testing set, wherein the one screenshot that overlapped had a different location and opacity value for the touch indicator.} During testing, we found that a training set of 15k screens was large enough to train the model to extremely high levels of accuracy (\ie > 97\%). 
To train the model we use the \textit{TensorFlow Object Detection API}~\cite{TFODA} that provides functions and configurations of well-known DL architectures. We provide details regarding our training process for \vtss object detection network in \secref{subsec:study-rq1}. Note that, despite the training procedure being completely automated, it needs to be run only once for a given device screen size, after which it can be re-used for inference. After the model is trained, inference is run on each frame, resulting in a set of output \textit{bounding box} predictions for each screen, with a confidence score.

\subsubsection{Opacity CNN}
Once \vts has localized the screen touches that exist in each video frame, it must then determine the opacity of each detected touch indicator to aid in the \textit{Action Classification} phase. This will help \vts in identifying instances where there are multiple actions in consecutive frames with very similar locations (\eg double tapping). To differentiate between low and high-opacity touch indicators, \vts adopts a modified version of the \AlexNet \cite{Krizhevsky:NIPS12} architecture as an {\sc Opacity CNN} that predicts whether a cropped image of the touch indicator is fully opaque (\ie finger touching screen) or low opacity (\ie indicating a finger being lifted off the screen). Similar to the object detection network, we fully automate the generation of the training dataset and training process for practicality. We again make use of the \ReDraw dataset and randomly select 10k unique screenshots, randomly crop a region of the screenshot to the size of a touch indicator, and an equal number of full and partial opacity examples are generated. For the low-opacity examples, we varied the transparency levels between 20\%-80\% to increase the diversity of samples in the training set. During initial experiments, we found that our {\sc Opacity CNN} required fewer training samples than the object detection network to achieve a high accuracy (\ie > 97\%). \NEW{Similar to the} \Faster model, this is a one-time training process, however, this model can be re-used across varying screen dimensions. Finally, \vts runs the classification for all the detected touch indicators found in the previous step. Then  \vts generates a \texttt{JSON} file containing all the detected bounding boxes, confidence levels, and opacity classifications. 

\subsection{Phase 2: Action Classification}
\label{subsec:approach-classification}

The \texttt{JSON} file generated by the \textit{Touch Detection} phase contains detailed data about the bounding boxes, opacity information, and the frame of each detected touch indicator (note we use the term ``touch indicator'' and ``touch'' interchangeably moving forward). This \texttt{JSON} file is used as input into the \textit{Action Classification} phase where single touches are grouped and classified as high-level actions. The classification of these actions involves two main parts: (i) a \textit{grouping algorithm} that associates touches across subsequent frames as a discrete action; and (ii) \textit{action translation}, which identifies the grouped touches as a single action type. The output of this step is a series of touch groups, each corresponding to an action type: (i) \texttt{Tap}, (ii) \texttt{Long Tap}, or (iii) \texttt{Gesture} (\eg swipes, pinches, etc).

\begin{figure}[t]
    \begin{center}
		\includegraphics[width=0.8\columnwidth]{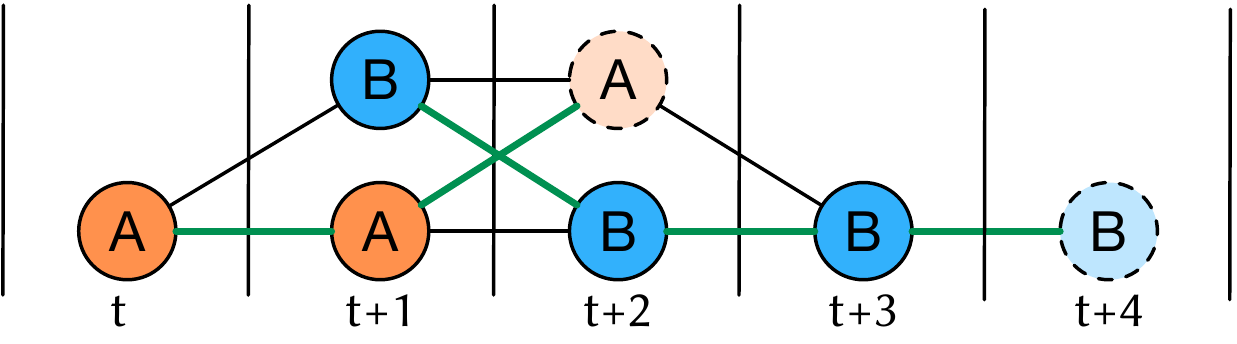}
		\vspace{-0.3cm}
        \caption{Illustration of the graph traversal problem for splitting discrete actions. Faded nodes with dotted lines represent touches where a finger is being lifted off the screen.}
        \label{fig:split_groups}
        \vspace{-0.6cm}
    \end{center}
\end{figure}

\subsubsection{Action Grouping}
\label{subsubsec:approach-grouping}

The first step of \vtss \textit{action grouping} filters out detected touches where the model's confidence is lower than 0.7. The second step groups touches belonging to the same atomic action according to a tailored heuristic and a graph connection algorithm. This procedure is necessary because discrete actions performed on the screen will persist across several frames, and thus, need to be grouped and segmented accordingly.

\noindent\textbf{Grouping Consecutive Touches.} The first heuristic groups touch indicators present in consecutive frames into the same group. As a measure taken to avoid (the rare occurrence of) a falsely detected touch indicator, touches that exist across two or fewer frames are discarded. This is due to the fact that, we observed in practice, even the quickest of touchscreen taps last across at least five frames. 

\noindent\textbf{Discrete Action Segmentation.} There may exist successive touches that were carried out extremely fast, such that there is no empty frame between two touches. In other cases, the touch indicator of one action may not fully disappear before the user starts a new action, leading to two or more touch indicators appearing in the same frame. These two situations are common when a user is swiping a list and quickly tapping an option, or typing quickly on the keyboard, respectively. However, it can be hard to determine where one action ends and another begins. 

\vts analyzes groups of consecutive or overlapping touches and segments them into discrete actions using a heuristic-based approach. We model the grouping of touch indicators as a graph connectivity problem (see  \figref{fig:split_groups}). In this formulation, touch indicators are represented as nodes, vertical lines separate consecutive frames, and edges are possible connections between consecutive touches that make up a discrete action. The goal is to derive the proper edges for traversing the graph such all touches for action $A$ are in one group and all touches for action $B$ are in another group (illustrated in green in \figref{fig:split_groups}). Our algorithm decomposes the lists of consecutive touches grouped together into a graph. Starting from the first node, our algorithm visits each subsequent node and attempts to link it to the previous node. If there is only one node in a subsequent frame, then two successive nodes are linked. If there is more than one node in a subsequent frame, our algorithm looks at the spatial distance between the previous node and both subsequent nodes, and groups previous nodes to their closer neighbors (as shown between frame $t$ and $t+1$). However, if multiple nodes in one frame are at similar distance from the previous node (as between frames $t+1$ and $t+2$ in \figref{fig:split_groups}), then the opacity of the nodes is used to perform the connection. For example, it is clear that node $A$ in frame $t+2$ is a finger being lifted off the screen. Thus, it must be connected to the previously occurring action $A$ (\ie in $t+1$), and not the action $B$ that just started.

Finally, after this process, the opacity of all linked nodes are analyzed to determine further splits. There exist multiple actions with no empty frame between them. Therefore, if a low-opacity node is detected in a sequence of successively connected nodes, they are split into distinct groups representing different actions.

\subsubsection{Action Translation}
\label{subsubsec:approach-translation}

This process analyzes the derived groups of touches and classifies them based on the number and touch locations in each group. For touches that start and end in the same spatial location on the screen (\eg the center of subsequent touch indicator bounding boxes varies by less than 20 pixels) the action is classified as a \texttt{Tap} or a \texttt{Long Tap}. \texttt{Taps} are recognized if the action lasts across 20 or fewer frames, and \texttt{Long-Taps} otherwise. Everything else is classified as a \texttt{Gesture}.

\noindent\textbf{Filtering.}
\vts removes actions that have touch indicators with low average opacity (\eg < 0.1\%) across a group, as this could represent a rare \textit{series} of misclassified touch indicators from the \Faster. \vts also removes groups whose size is below or equals a threshold of two frames, as these might also indicate rare misclassifications. The result of the \textit{Action Classification} phase is a structured list of actions (\ie \texttt{Tap}, \texttt{Long Tap}, or \texttt{Gesture}), where each action is a series of touches associated to video frames and screen locations.

\vspace{-0.1cm}
\subsection{Phase 3: Scenario Generation}
\label{subsec:approach-generation}

After all the actions have been derived by the \textit{Action Classification} phase, \vts proceeds by generating commands using the Android Debug Bridge (\texttt{adb}) that replay the classified actions on a device. To accomplish this, \vts converts the classified, high-level actions into low-level instructions in the \texttt{sendevent} command format, which is a utility included in Android's Linux kernel. Then, \vts uses a modified RERAN~\cite{Gomez:ICSE13} binary to replay the events on a device.

\noindent\textbf{Generating the Scenario Script.} The \texttt{sendevent} command uses a very limited instruction set in order to control the UI of an Android device. The main instructions of interest are the \texttt{start\_event}, \texttt{end\_event}, $x$ and $y$ coordinates where the user's finger touched the screen, and certain special instructions required by devices with older API levels. To create the script, each action is exported starting with the \texttt{start\_event} command. Then, for actions classified as a \texttt{Tap}, \vts provides a single $(x, y)$ coordinate pair, derived from the center of the detected bounding box of the \texttt{Tap}. For \texttt{Gestures}, \vts iterates over each touch that makes up the \texttt{Gesture} action and appends the  $(x, y)$ pairs of each touch indicator to the list of instructions. For \texttt{Long Taps}, \vts performs similar processing to that of \texttt{Gestures}, but instead uses only a single  $(x, y)$ pair from the initial detected touch indicator bounding box.

Then, \vts ends the set of instructions for an action with the appropriate \texttt{end\_event} command. For \texttt{Gestures} and \texttt{Long Taps} the speed and duration of each instruction is extremely important in order to accurately replay the user's actions. To derive the speed and duration of these actions, \vts adds timestamps to each  $(x, y)$ touch location based on the timing between video frames (\ie for 30fps, there is a 33 millisecond delay between each frame), which will temporally separate each touch command sent to the device.  
The same concept applies for \texttt{Long Tap}, however, since this action type uses a single  $(x, y)$ touch location, the timing affects the duration that the touch event lasts on the screen. 

Finally, in order to determine the delays between successive actions, the timing between video frames is again used. Our required 30fps frame-rate provides \vts with millisecond-level resolution of event timings, whereas higher frame-rates will only increase the fidelity of replay timing. 

\noindent\textbf{Scenario Replay.} Once all the actions have been converted into low-level \texttt{sendevent} instructions, they are written to a log file. This log file is then fed into a translator which converts the file into a runnable format that can be directly replayed on a device. This converted file along with a modified version of the RERAN engine~\cite{Gomez:ICSE13} is pushed to the target device. We optimized the original RERAN binary to replay event traces more efficiently. Finally, the binary is executed using the converted file to faithfully replay the user actions originally recorded in the initial input video. We provide examples of \vts's generated \texttt{sendevent} scripts, alongside our updated version of the RERAN binary in our online appendix~\cite{appendix}.

\vspace{-0.1cm}
\section{Design of the Experiments}
\label{sec:study}
In this section, we describe the procedure we used to evaluate \vts. The goal of our empirical study is to assess the \textit{accuracy}, \textit{robustness}, \textit{performance}, and \textit{industrial utility} of the approach. 
The \textit{context} of this evaluation consists of: (i) sets of 15,000 and 10,000 images, corresponding to the evaluation of \approachs \Faster and {\sc Opacity CNN} respectively; (ii) a set of 83 Android applications including 68 of the top-rated apps from Google Play, five open source apps with real crashes, five open source apps with known bugs, and five open source apps with controlled crashes; (iii) two popular target Android devices (the Nexus 5 and Nexus 6P)\footnote{\NEW{It should be noted that \approach can be used with emulators via minor modifications to the script generation process}}. The main \textit{quality focus} of our study is the extent to which \vts can generate replayable scenarios that mimic original user GUI inputs. To achieve our study goals, we formulated the following five research questions:

\begin{itemize}
	\item{\textit{\textbf{RQ$_1$}: How accurate is \approach in identifying the location of the touch indicator?}} 
	\item{\textit{\textbf{RQ$_2$}: How accurate is \approach in identifying the opacity of the touch indicator?}}
	\item{\textit{\textbf{RQ$_3$}: How effective is \approach in generating a sequence of events that accurately mimics the user behavior from video recordings of different applications?}}
	\item{\textit{\textbf{RQ$_4$}: What is \approach's overhead in terms of scenario generation?}}
	\item{\textit{\textbf{RQ$_5$}: Do practitioners perceive \vts as useful?}}
\end{itemize}

\subsection{RQ$_1$: Accuracy of Faster R-CNN}
\label{subsec:study-rq1}

To answer RQ$_1$, we first evaluated the ability of \vtss \Faster to accurately identify and localize the \textit{touch indicators} present in screen recording frames with \textit{bounding boxes}. To accomplish this, we followed the procedure to generate training data outlined in \secref{subsec:approach-detection} complete with the 70\%--30\% split for the training and testing sets, respectively. The implementation of the \Faster object detection used by \vts is coupled to the size of the images used for training and inference. Thus, to ensure \vtss model functions across different devices, we trained two separate \Faster models (\ie one for the Nexus 5 and one for the Nexus 6P), by resizing the images from the \ReDraw dataset to the target device image size. As we show in the course of answering other RQs, we found that resizing the images in the already large \ReDraw dataset, as opposed to re-collecting natively sized images for each device, resulted in highly accurate detections in practice. 

We used the \textit{TensorFlow Object Detection API}~\cite{TFODA} to train our model. Moreover, for the training process, we modified several of the hyper-parameters after conducting an initial set of experiments.  These changes affected the number of classes (\ie 1), maximum number of detections per class or image (\ie 10), and the learning rate after 50k (\ie $3\times10^{-5}$) and 100k (\ie $3\times10^{-6}$) iterations. The training process was run for 150k steps with a batch size of 1, and our implementation of \Faster utilized a \VGGNet~\cite{Simonyan:ICLR14} instance pre-trained on the MSCOCO dataset~\cite{Lin:LNCS14}. We provide our full set of model parameters in our online appendix~\cite{appendix}.

To validate the accuracy of \vtss \Faster models we utilize \textit{Mean Average Precision} (mAP) which is commonly used to evaluate techniques for the object detection task. This metric is typically computed by averaging the \textit{precision} over all the categories in the data, however, given we have a single class (the touch indicator icon), we report results only for this class. Thus, our mAP is computed as $mAP=TP/(TP+FP)$ where $TP$ corresponds to an identified image region with a correct corresponding label, and $FP$ corresponds to the identified image regions with the incorrect label (which in our case would be an image region that is falsely identified as a touch indicator). Additionally, we evaluate the \textit{Average Recall} of our model in order to determine if our model misses detecting any instances of the touch indicator. This is computed \NEW{by $AR=TP/k$} where $TP$ is the same definition stated above, and \NEW{the $k$ corresponds to} the total number of possible $TP$ predictions.

\begin{figure}[t]
    \centering
    \begin{subfigure}[b]{0.4\linewidth}
        \centering
        \includegraphics[height=0.3in]{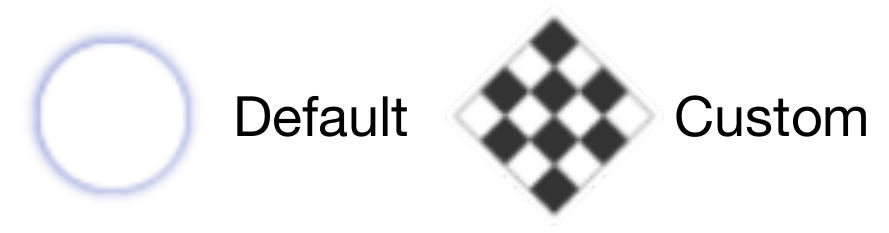}
        \caption{Touch indicators}
        \label{fig:touches}
        \vspace{-0.5em}
    \end{subfigure}%
     ~
    \begin{subfigure}[b]{0.5\linewidth}
        \centering
        \includegraphics[height=0.3in]{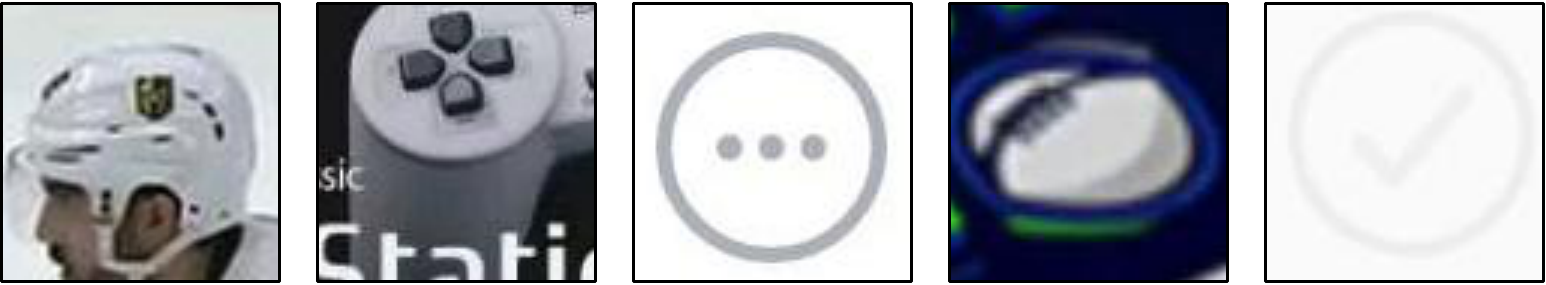}
        \caption{False positive detections}
        \label{fig:false_positives}
        \vspace{-0.5em}
    \end{subfigure}
    \caption{Touch indicators and failed detections}
    \label{fig:touches_false_positives}
     \vspace{-0.5cm}
\end{figure}

During preliminary experiments with \Faster using the default \textit{touch indicator} (see \figref{fig:touches}), we found that, due to the default touch indicator's likeness to other icons and images present in apps, it was prone to very occasional false positive detections (\figref{fig:false_positives}). Thus, we analyzed particular cases in which the default touch indicator failed and replaced it with a more distinct, high-contrast touch indicator. We found that this custom touch indicator marginally improved the accuracy of our models. It should be noted that replacing the touch indicator on a device, requires the device to be rooted. While this is an acceptable option for most developers, it may prove difficult for end-users. However, even with the default touch indicator, \vtss \Faster model still achieves extremely high levels of accuracy. 

\subsection{RQ$_2$: Accuracy of Opacity CNN} 
\label{subsection:study-rq2}

To answer RQ$_2$, we evaluated the ability of \vts's {\sc Opacity CNN} to predict whether the opacity of the touch indicator is solid or semi-transparent. To accomplish this, we followed dataset generation procedure outlined in \secref{subsec:approach-detection}, where equal number of full and partial opacity examples are generated.
Thus, the generated dataset contains equal numbers of full and partial opacity examples for a total of 10k, which are evenly split into 70\%--30\% training and testing sets. We used the TensorFlow framework in combination with Keras to implement the {\sc Opacity CNN}. In contrast to the \Faster model used previously, we do not need to create a separate model for each device. This is due to the \textit{touch indicator} being resized when fed into the {\sc Opacity CNN}. Similarly to the \Faster, we evaluate {\sc Opacity CNN} using \textit{mAP} across our two classes.

\subsection{RQ$_3$: Accuracy on Different Scenarios}
\label{subsec:study-rq3}

To answer RQ$_3$, we carried out two studies designed to assess both the \textit{depth}, and \textit{breadth} of \vtss abilities to reproduce user events depicted in screen recordings. The first, \textit{Controlled Study}, measures the depth of \vtss abilities through a user study during which we collected real videos \NEW{from end users depicting: bugs, real crashes, synthetically injected crashes, and normal usage scenarios for 20 apps}. Next in the \textit{Popular Applications Study} we measured the breadth of \vtss abilities by recording scenarios for a larger, more diverse set of 64 most popular apps from the Google Play. We provide the full details of these apps in our online appendix~\cite{appendix}. 

\subsubsection{Controlled Study}

In this study we, considered four types of recorded usage scenarios depicting: (i) normal usages, (ii) bugs, (iii) real crashes, and (iv) controlled crashes. Normal usage scenarios refer to video recordings exercising different features on popular apps. Bug scenarios refer to video recordings that exhibit a bug on open source apps. Finally, controlled crashes refer to injected crashes into open source apps. This allows us to control the number of steps before the crash is triggered. 

For this study, eight participants \NEW{including 1 undergraduate, 3 masters, and 4 doctoral students} were recruited from William \& Mary (approved by the Protection of Human Subjects Committee (PHSC) at W\&M under protocol PHSC-2019-01-22-13374) to record the videos, with each participant recording eight separate videos, two from each of the categories listed above. Four participants recorded videos on the Nexus 5 and four used the Nexus 6P. This accounts for a total of 64 videos, from 20 apps evenly distributed across all scenarios. Before recording each app, participants were either asked to use the app to become familiar with it, or read a bug/crash report before reproducing the fault. All of the real bugs and crashes were taken from established past studies on mobile testing and bug reporting~\cite{Moran:ICST16,Moran:FSE15,Chaparro:FSE'19}.

\subsubsection{Popular Applications Study}
For the next study, we considered a larger more diverse set of apps from Google Play. More specifically we downloaded the two highest-rated apps from each non-game category (\ie 32) for a total of 64 applications. 

Two of the authors then recorded two scenarios per app accounting for 32 apps each, one using the Nexus 5 and the other using a Nexus 6P. The authors strived to use the apps as naturally as possible, and this evaluation procedure is in line with past evaluations of Android record-and-replay techniques~\cite{Qin:ICSE16,Gomez:ICSE13}. The recorded scenarios represented specific use cases of the apps that exercise at least one of the major features, and were independent of one another. During our experiments, we noticed certain instances where our recorded scenarios were not replicable, either due to non-determinism or dynamic content (\eg random popup appearing). Thus, we discarded these instances and were left with 111 app usage scenarios from 60 apps. \NEW{It is worth noting that it would be nearly impossible for existing techniques such as RERAN~\cite{Gomez:ICSE13} or Barista~\cite{Fazzini:ICST17} to reproduce the scenarios due to the nondeterminism of the dynamic content, hence our decision to exclude them.
}

To measure how accurately \vts replays videos, we use three different metrics. To compute these metrics, we manually derived the ground truth sequence of action types for each recorded video. First, we use Levenshtein distance, which is commonly used to compute distances between words at character level, to compare the original list of action types to the list of classified actions generated by \vts. Thus, we consider each type of action being represented as a character, and scenarios as sequences of characters which represent series of actions. A low Levenshtein distance value indicates fewer required changes to transform \approachs output to the ground truth set of actions. Additionally, we compute the longest common subsequence (LCS) to find the largest sequence of each scenario from \vtss output that aligns with the ground truth scenario from the video recording. For this LCS measure, the higher the percentage, the closer \approachs trace is to a perfect match of the original trace. Moreover, we also computed the precision and recall for \vts to predict each type of action across all scenarios when compared to the ground truth. Finally, in order to validate the fidelity of the replayed scenarios generated by \vts compared to the original video recording, we manually compared each original video to each reproduced scenario from \vts, and determined the number of actions for each video that were faithfully replayed.

\subsection{RQ$_4$: Performance}
\label{subsec:study-rq4}

To investigate RQ$_4$, we evaluated \vts by calculating the average time it takes for a video to pass through each of the three phases of the \vts approach on commodity hardware (\ie a single NVIDIA GTX 1080Ti). We see this as a worst case scenario for \vts performance, as our approach could perform substantially faster on specialized hardware. Note that since our replay engine is an enhancement of the RERAN engine, we expect our scripts to have similar or better overhead as reported in its respective paper \cite{Gomez:ICSE13}.

\subsection{RQ$_5$: Perceived Usefulness}

Ultimately, our goal is to integrate \vts into real-world development environments. Thus, as part of our evaluation, we investigated \vtss perceived usefulness with three developers who build Android apps (or web apps for mobile) for their respective companies.

The developers (\aka participants) were contacted through direct contact of the authors. Participant~\#1 (P1) was a front-end developer on the image search team of the Google Search app~\cite{g-image-search}, participant~\#2 (P2) is a developer of the 7-Eleven Android app~\cite{7-eleven}, and participant~\#3 (P3) is a backend developer for the Proximus shopping basket app~\cite{proximus}. We interviewed the participants using a set of questions organized in two sections. The first section aimed to collect information on participants' background, including their role at the company, the information used to complete their tasks, the quality of this information, the challenges of collecting it, and how they use videos in their every-day activities. The second section aimed to assess \vtss potential usefulness as well as its accuracy in generating replayable scenarios. This section also asked the participants for feedback to improve \vts including likert scale questions. We provide the complete list of interview questions used in our online appendix ~\cite{appendix}, and discuss selected questions in \secref{subsec:results-rq5}.

The participants answered questions from the second section by comparing two videos showing the same usage scenario for their respective app: one video displaying the scenario manually executed on the app, and the other one displaying the scenario executed automatically via \vtss generated script. Specifically, we defined, recorded, and manually executed a usage scenario on each app. Then, we ran \vts on the resulting video recordings. To define the scenarios, we identified a feature on each app involving any of the action types (\ie taps, long taps, and gestures). Then, we generated a video showing the original scenario (\ie video recording) and next to it the replayed scenario generated when executing \vtss script. Both recordings highlight the actions performed on the app. We presented the video to participants as well as \vtss script with the high-level actions automatically identified from the original video.

\section{Empirical Results}
\label{sec:results}

\subsection{RQ$_1$: Accuracy of \Faster}
\label{subsec:results-rq1}

\begin{table}[tb]
	\footnotesize
	\centering
	\caption{Touch Indicator Detection Accuracy}
	\vspace{-0.9em}
	\label{tab:rq1_map_ar}
	\begin{tabular}{|l|c|c|c|c|} \hline
		\textbf{Model}&\textbf{Device}&\textbf{mAP}&\textbf{mAP@.75}&\textbf{AR}\\
		\hline
		\Faster-Original&Nexus 5&97.36\%&99.01\%&98.57\%\\
		\Faster-Original&Nexus 6P&96.94\%&99.01\%&98.19\%\\
		\Faster-Modified&Nexus 5&97.98\%&99.01\%&99.33\%\\
		\Faster-Modified&Nexus 6P&97.49\%&99.01\%&99.07\%\\
		\hline
	\end{tabular}
	\vspace{-0.3cm}
\end{table}

\tabref{tab:rq1_map_ar} depicts the precision and recall for \vtss \Faster network for touch indicator detection on different devices and datasets. The first column identifies the usage of either the default touch indicator or the modified version. The second column describes the target device for each trained model. The third column provides the mAP regardless of the Intersection Over Union (IoU) \cite{Ren:NIPS15} between the area of the prediction and the ground truth. \NEW{The forth column presents the AP giving the proportion of $TP$ out of the possible positives.} All models achieve $\approx$97\% mAP, indicating that \vtss object detection network is highly accurate. The mAP only improves when we consider bounding box IoUs that match the ground truth bounding boxes by at least 75\%, which illustrates that when the model is able to predict a reasonably accurate bounding box, it nearly always properly detects the touch indicator ($\approx$99\%). As illustrated by the last column in \tabref{tab:rq1_map_ar}, the model also achieves extremely high recall, detecting at least $\approx$98\% of the inserted touch indicators.
\vspace{0.2cm}
\begin{tcolorbox}[enhanced,skin=enhancedmiddle,borderline={1mm}{0mm}{MidnightBlue}]
\textbf{Answer to RQ$_1$}: \vts benefits from the strong performance of its object detection technique to detect touch indicators. All \Faster models achieved at least $\approx$ 97\% precision \NEW{and at least $\approx$ 98\% recall} across devices.
\end{tcolorbox}  

\begin{table}[tb]
	\footnotesize
	\centering
	\caption{Confusion Matrix for Opacity CNN. Low Opacity Original (L-Op.-Orig.), High Opacity Original (H-Op.-Orig.), Low Opacity Custom (L-Op.-Cust.), High Opacity Custom (H-Op.-Cust.) }
	\vspace{-0.9em}
	\label{tab:rq2_cm_opacity}
	\begin{tabular}{|r|r|r|r|r|r|} \hline
		&Total&L-Op.-Orig.& H-Op.-Orig.&L-Op.-Cust.&H-Op.-Cust.\\
		\hline
		Low Op&5000&\cellcolor{MidnightBlue!20}97.8\%&2.2\%&\cellcolor{MidnightBlue!20}99.7\%&0.3\%\\
		High Op&5000&1.4\%&\cellcolor{MidnightBlue!20}98.6\%&0.8\%&\cellcolor{MidnightBlue!20}99.2\%\\
		\hline
	\end{tabular}
	\vspace{-0.3cm}
\end{table}

\subsection{RQ$_2$: Accuracy of the {\sc Opacity CNN}}
\label{subsec:results-rq2}

To illustrate the {\sc Opacity Network}'s accuracy in classifying the two opacity levels of touch indicators, we present the confusion matrix in \tabref{tab:rq2_cm_opacity}. The results are presented for both the default and modified touch indicator. The overall top-1 precision for the original touch indicator is 98.2\% whereas for the custom touch indicator is 99.4\%. \NEW{These percentages are computed by aggregating the correct identifications for both classes (i.e., Low/High-Opacity) \textit{together} for the original and custom touch indicators.} Hence, it is clear \vtss~{\sc Opacity CNN} is highly effective at distinguishing between differing opacity levels.
\begin{tcolorbox}[enhanced,skin=enhancedmiddle,borderline={1mm}{0mm}{MidnightBlue}]
\textbf{Answer to RQ$_2$}: \vts benefits from the \Cnns accuracy in classifying levels of opacity. {\sc Opacity CNN} achieved an average precision above 98\% for both touch indicators.
\end{tcolorbox}  

\begin{table*}[]
\footnotesize
\caption{Detailed Results for RQ$_3$ popular applications study. Green cells indicate fully reproduced videos, orange cells >50\% reproduced, and Red Cells <50\% reproduced. Blue cells show non-reproduced videos due to non-determinism/dynamic content.}
\vspace{-1.2em}
\label{tab:rq3-all-results}
\begin{tabular}{|l|
>{\columncolor[HTML]{9AFF99}}l |
>{\columncolor[HTML]{9AFF99}}l |l|
>{\columncolor[HTML]{9AFF99}}l |l|l|
>{\columncolor[HTML]{9AFF99}}l |
>{\columncolor[HTML]{9AFF99}}l |l|
>{\columncolor[HTML]{9AFF99}}l |
>{\columncolor[HTML]{9AFF99}}l |l|
>{\columncolor[HTML]{9AFF99}}l |
>{\columncolor[HTML]{9AFF99}}l |}
\hline
\cellcolor[HTML]{C0C0C0}\textbf{AppName} & \multicolumn{2}{l|}{\cellcolor[HTML]{C0C0C0}\textbf{Rep. Actions}} & \cellcolor[HTML]{C0C0C0}\textbf{AppName} & \multicolumn{2}{l|}{\cellcolor[HTML]{C0C0C0}\textbf{Rep. Actions}} & \cellcolor[HTML]{C0C0C0}\textbf{App Name} & \multicolumn{2}{l|}{\cellcolor[HTML]{C0C0C0}\textbf{Rep. Actions}} & \cellcolor[HTML]{C0C0C0}\textbf{App Name} & \multicolumn{2}{l|}{\cellcolor[HTML]{C0C0C0}\textbf{Rep. Actions}} & \cellcolor[HTML]{C0C0C0}\textbf{App Name} & \multicolumn{2}{l|}{\cellcolor[HTML]{C0C0C0}\textbf{Rep. Actions}} \\ \hline
Ibis Paint X                    & 11/11                           & 36/36                             & Firefox                         & 22/22                            & \cellcolor[HTML]{D9E0F2}N/A      & Tasty                            & 20/20                            & \cellcolor[HTML]{FFCB2F}14/36    & SoundCloud                       & 12/12                            & 13/13                            & LetGo                            & 17/17                             & 15/15                           \\ \hline
Pixel Art Pro                   & 20/20                           & 9/9                               & MarcoPolo                       & 12/12                            & \cellcolor[HTML]{9AFF99}30/30    & Postmates                        & 26/26                            & 12/12                            & Shazam                           & \cellcolor[HTML]{FFCB2F}12/15    & 20/20                            & TikTok                           & 14/14                             & 11/11                           \\ \hline
Car-Part.com                    & 20/20                           & 16/16                             & Dig                             & 13/13                            & \cellcolor[HTML]{D9E0F2}N/A      & Calm                             & 9/9                              & \cellcolor[HTML]{FFCB2F}11/16    & Twitter                          & 14/14                            & 19/19                            & LinkedIn                         & 18/18                             & 13/13                           \\ \hline
CDL Practice                    & 8/8                             & 13/13                             & Clover                          & 15/15                            & \cellcolor[HTML]{9AFF99}19/19    & Lose It!                         & 36/36                            & \cellcolor[HTML]{D9E0F2}N/A      & News Break                       & \cellcolor[HTML]{F56B00}1/18     & 9/9                              & CBSSports                        & 25/25                             & 16/16                           \\ \hline
Sephora                         & \cellcolor[HTML]{FFCC67}4/9     & 13/13                             & PlantSnap                       & 39/39                            & \cellcolor[HTML]{FFCB2F}18/24    & U Remote                 & 14/14                            & 18/18                            & FamAlbum                      & 19/19                            & \cellcolor[HTML]{FFCB2F}8/26     & MLBatBat                         & \cellcolor[HTML]{FFCB2F}11/13     & \cellcolor[HTML]{D9E0F2}N/A     \\ \hline
SceneLook                       & 14/14                           & 16/16                             & Translator                      & 20/20                            & \cellcolor[HTML]{9AFF99}28/28    & LEGO                       & 52/52                            & 24/24                            & Baby-Track                     & 14/14                            & 12/12                            & G-Translate                 & \cellcolor[HTML]{FFCB2F}14/17     & 15/15                           \\ \hline
KJ Bible                        & 16/16                           & 19/19                             & Tubi                            & \cellcolor[HTML]{FD6864}2/19     & \cellcolor[HTML]{9AFF99}30/30    & Dev Libs                         & 35/35                            & 22/22                            & Walli                            & 8/8                              & \cellcolor[HTML]{D9E0F2}N/A      & G-Podcast                  & 9/9                               & 15/15                           \\ \hline
Bible App                       & 12/12                           & 15/15                             & Scan Radio                   & \cellcolor[HTML]{FFCC67}24/27    & \cellcolor[HTML]{D9E0F2}N/A      & Horoscope                        & 24/24                            & 19/19                            & ZEDGE                            & 9/9                              & 18/18                            & Airbnb                           & 9/9                               & 14/14                           \\ \hline
Indeed Jobs                     & 15/15                           & 19/19                             & Tktmaster                    & 30/30                            & \cellcolor[HTML]{9AFF99}14/14    & Waze                             & 17/17                            & 19/19                            & G-Photo                    & 18/18                            & 18/18                            & G-Earth                     & \cellcolor[HTML]{F56B00}8/13      & \cellcolor[HTML]{F56B00}1/30    \\ \hline
UPS Mobile                      & 16/16                           & \cellcolor[HTML]{FFCB2F}19/24     & Greet Cards                   & 23/23                            & \cellcolor[HTML]{D9E0F2}N/A      & Transit                          & 26/26                            & 18/18                            & PicsArt                          & 18/18                            & 39/39                            & DU Record                      & 15/15                             & 9/9                             \\ \hline
Webtoon                         & 17/17                           & \cellcolor[HTML]{FFCB2F}15/21     & QuickBooks                      & 47/47                            & \cellcolor[HTML]{9AFF99}28/28    & WebMD                            & \cellcolor[HTML]{FD6864}7/34     & \cellcolor[HTML]{FD6864}7/26     & G-Docs                      & \cellcolor[HTML]{F56B00}3/26     & \cellcolor[HTML]{D9E0F2}N/A      & AccuWeath                      & 13/13                             & 21/21                           \\ \hline
MangaToon                       & 16/16                           & 28/28                             & Yahoo Fin                   & 23/23                            & \cellcolor[HTML]{D9E0F2}N/A      & K-Health                         & 10/10                            & \cellcolor[HTML]{FFCB2F}15/24    & M. Outlook                     & 27/27                            & \cellcolor[HTML]{FFCB2F}21/26    & W. Radar                    & 14/14                             & 13/13                           \\ \hline
\end{tabular}
\end{table*}

\vspace{-0.2cm}
\subsection{RQ$_3$: Scenario Replay Accuracy}
\label{subsec:results-rq3}

\begin{figure}[t]
    \centering
    \begin{subfigure}[b]{0.25\linewidth}
        \centering
        \includegraphics[height=0.8in]{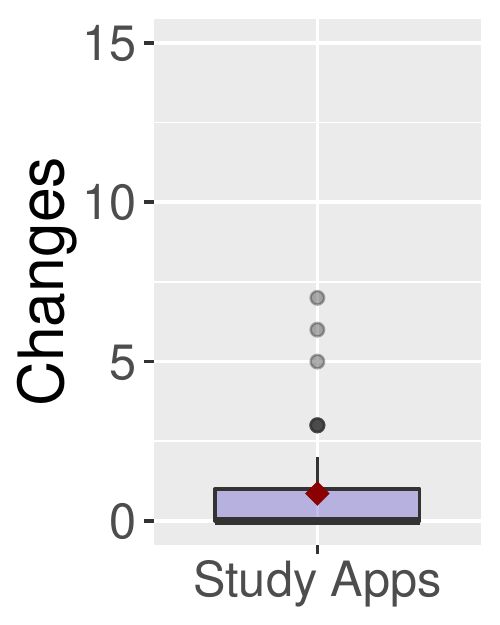}
        \caption{LD-Study}
        \label{fig:rq3_ld}
        \vspace{-0.5em}
    \end{subfigure}%
     ~
    \begin{subfigure}[b]{0.25\linewidth}
        \centering
        \includegraphics[height=0.8in]{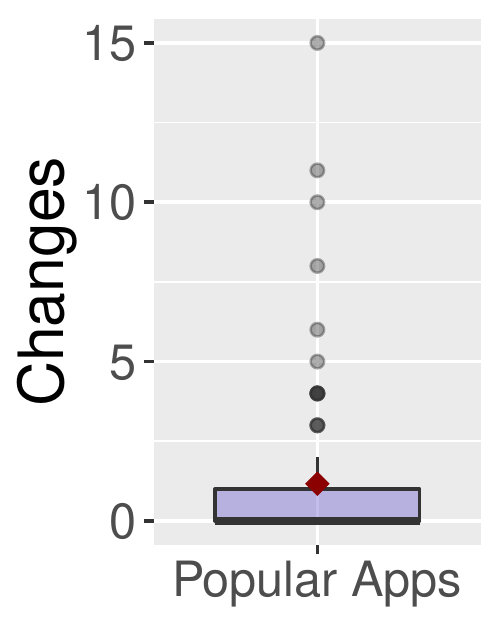}
        \caption{LD-Popular}
        \label{fig:rq3_ld_p}
        \vspace{-0.5em}
    \end{subfigure}%
    ~ 
    \begin{subfigure}[b]{0.25\linewidth}
        \centering
        \includegraphics[height=0.8in]{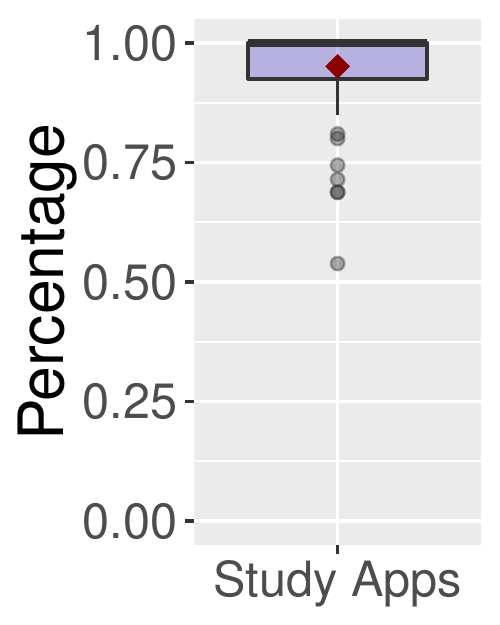}
        \caption{LCS-Study}
        \label{fig:rq3_lcs}
        \vspace{-0.5em}
    \end{subfigure}
    ~ 
    \begin{subfigure}[b]{0.25\linewidth}
        \centering
        \includegraphics[height=0.8in]{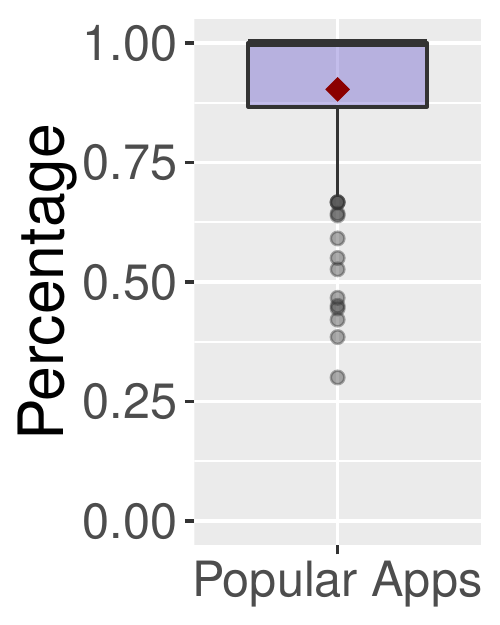}
        \caption{LCS-Popular}
        \label{fig:rq3_lcs_p}
        \vspace{-0.5em}
    \end{subfigure}
    \caption{Effectiveness Metrics}
    \label{fig:rq3_ld_lcs}
\vspace{-0.3cm}
\end{figure}

\noindent \textbf{Levenshtein Distance.} \figref{fig:rq3_ld} and \ref{fig:rq3_ld_p} depict the number of changes required to transform the output event trace into the ground truth for the apps used in the \textit{controlled study} and the \textit{popular apps} study, respectively. For the controlled study apps, on average it requires $0.85$ changes per user trace to transform \vtss output into ground truth event trace, whereas for the popular apps it requires slightly more with $1.17$ changes. Overall, \vts requires minimal changes per event trace, being very similar to the ground truth. 

\noindent\textbf{Longest Common Subsequence.} \figref{fig:rq3_lcs} and \ref{fig:rq3_lcs_p} presents the percentage of events for each trace that match those in the original recording trace for the \textit{controlled study} and \textit{popular apps} study respectively. On average \vts is able to correctly match 95.1\% of sequential events on the ground truth for the controlled study apps and 90.2\% for popular apps. These results suggest that \vts is able to generate sequences of actions that closely match the original trace in terms of action types. 

\begin{figure}[t]
    \begin{center}
		\includegraphics[width=\linewidth]{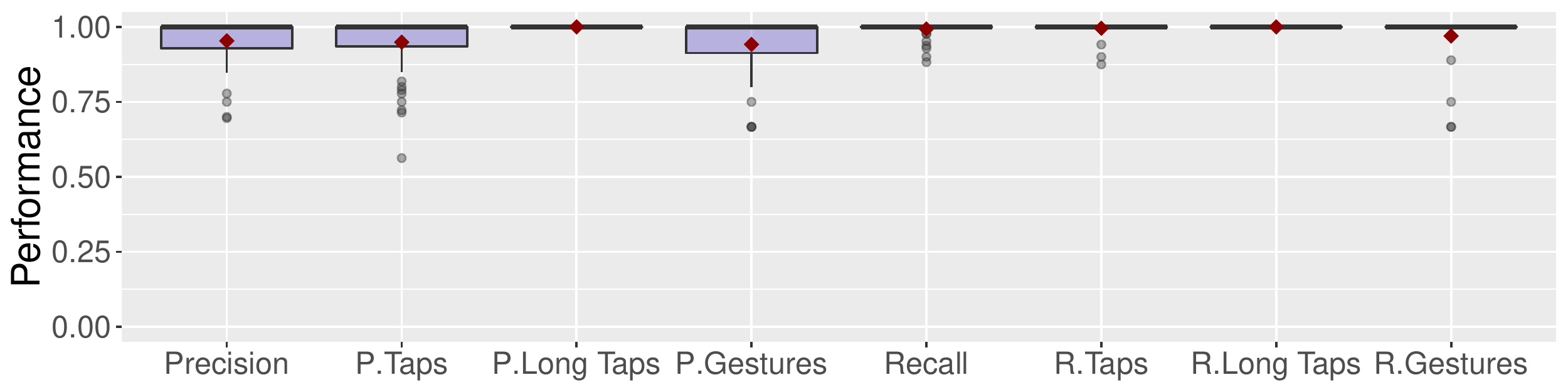}
		\vspace{-0.65cm}
        \caption{Precision and Recall - Controlled Study}
        \label{fig:rq3_precision_recall}
    \end{center}
\vspace{-0.2cm}
\end{figure}

\begin{figure}[t]
    \begin{center}
		\includegraphics[width=\linewidth]{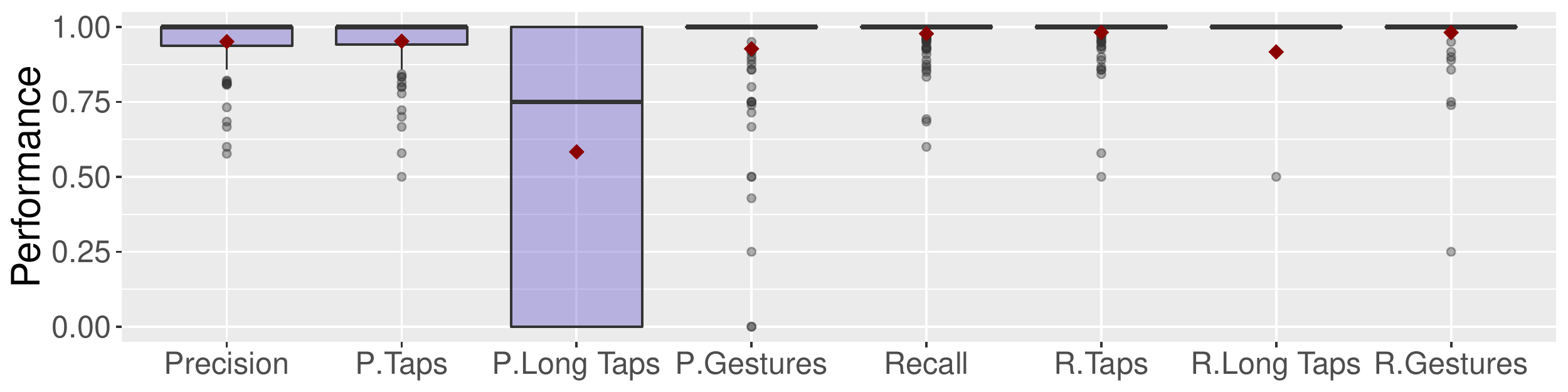}
		\vspace{-0.65cm}
        \caption{Precision and Recall - Popular Apps}
        \label{fig:rq3_precision_recall_p}
    \end{center}
\vspace{-0.5cm}
\end{figure}

\noindent\textbf{Precision and Recall.} \figref{fig:rq3_precision_recall} and \ref{fig:rq3_precision_recall_p} show the precision and recall results for the \textit{controlled study} and \textit{popular apps study}, respectively. These plots were constructed by creating an order agnostic ``bag of actions'' for each predicted action type, for each scenario in our datasets. Then the precision and recall are calculated by comparing the actions to a ground truth ``bag of actions'' to compute precision and recall metrics. Finally, an overall average precision and recall are calculated across all action types. The results indicate that on average, the precision of the event traces is 95.3\% for the controlled study apps and 95\% for popular apps. This is also supported for each type of event showing also a high level of precision across types except for the precision on \texttt{Long Taps} for the popular apps. This is mainly due to the small number (\ie 9 \texttt{Long Taps}) of this event type across all the popular app traces. Also, \figref{fig:rq3_precision_recall}  and \ref{fig:rq3_precision_recall_p} illustrate that the recall across action types is high with an average of 99.3\% on controlled study apps and 97.8\% on the popular apps for all types of events. In general, we conclude that \vts can accurately predict the correct number of event types across traces.

\noindent\textbf{Success Rate.} Finally, we also evaluated success rate of each replayed action for all scenarios across both RQ$_3$ studies. The 175 videos were analyzed manually and each action was marked as successful if the replayable scenario faithfully exercised the app features according to the original video. This means that in certain cases videos will not \textit{exactly} match the original video recording (\eg due to a single keyboard keystroke error that still led to the same feature result). Thus, after validating all 64 videos for the controlled study, \vts fully reproduces 93.75\% of the scenarios, and 94.48\% of the consecutive actions. \vts fully reproduced 96.67\% of the scenarios for bugs and crashes and 91.18\% of apps usages. Detailed results for the \textit{popular apps study} are shown in \tabref{tab:rq3-all-results}, where each app, scenario (with total number of actions), and successfully replayed actions are displayed. \NEW{\textit{Green} cells indicate a fully reproduced video, \textit{Orange} cells indicate more than 50\% of events reproduced, and Red cells indicate less than 50\% of reproduced events. \textit{Blue} cells show non-reproduced videos due to non-determinism/dynamic content.} For the 111 scenarios recorded for the popular apps, \vts fully reproduced 81.98\% scenarios, and 89\% of the consecutive actions. Overall, this signals strong replay-ability performance across a highly diverse set of applications. Instances where \vts failed to reproduce scenarios are largely due to minor inaccuracies in \texttt{Gesture} events due to our video resolution of 30fps. We discuss potential solutions to this limitation in \secref{sec:limitations}.

\vspace{0.1cm}
\begin{tcolorbox}[enhanced,skin=enhancedmiddle,borderline={1mm}{0mm}{MidnightBlue}]
\textbf{Answer to RQ$_3$}: \vts is capable of generating event traces that require on average $\approx$ 1 change to match original user scenarios. Moreover,  at least 90.2\% of events match the ground truth, when considering the sequence of event types. Overall, precision and recall are $\approx$95\% and $\approx$98\% respectively for event types produced by \vts. Finally,  in 96.67\%  and 91.18\% of the cases, \vts successfully reproduces  bugs/crashes- and app-usage-related videos, respectively.
\end{tcolorbox}

\subsection{RQ$_4$: Approach Performance}
\label{subsec:results-rq4}

To measure the performance of \vts, we measured the average time in seconds/frame (s/f) for a single video frame to be processed across all recorded videos for three components: (i) the frame extraction process (0.045 s/f), (ii) the touch indicator detection process (1.09 s/f), and (iii) the opacity classification process (0.032 s/f). The script generation time is negligible compared to these other processing times, and is only performed once per video. This means that an average video around 3 mins in length would take \vts $\approx$105 minutes to fully process and generate the script. However, this process is \textit{fully automated}, can run in the background, and can be accelerated by more advanced hardware. We expect the overhead of our replayed scripts to be similar or better than RERAN since \vts replay engine is essentially an improved version of RERAN's. 

\begin{tcolorbox}[enhanced,skin=enhancedmiddle,borderline={1mm}{0mm}{MidnightBlue}]
\textbf{Answer to RQ$_4$}: \approach is capable of fully processing an average 3-min screen recording in  $\approx$105 mins. 
\end{tcolorbox} 

\subsection{RQ$_5$: Perceived Usefulness}
\label{subsec:results-rq5}

The three industry participants agreed that further tool support is needed for helping QA members and other stakeholders with generating video recordings. For example,  P3 mentions that  while videos are "more useful than images" (in some cases), they ``may be difficult to record'' because of ``time constraints''. All participants also (strongly) agreed that the scenarios produced by \vts (in the generated scripts) are accurate with respect to the scenarios that were manually executed.

Regarding \vts's usefulness, P1 remarked that the QA team could use \vts to help them create videos more optimally. P2 supported this claim as he mentions that \vts could help ``the QA team write/provide commands or steps, then the tool would read and execute these while recording a video of the scenario and problem. This solution could be integrated in the continuous integration pipeline''. In addition, P3 mentions that \vts could be used during app demos: \vts could ``automatically execute a script that shows the app functionality and record a video. In this way, the demo would focus on business explanation rather than on manually providing input to the app or execute a user scenario''.

P2 also indicated that \vts could be used to analyze user behavior within the app, which can help improve certain app screens and navigation. He mentions that \vts ``could collect the type of interactions, \# of taps, \etc to detect, for example, if certain screens are frequently used or if users often go back after they go to a particular screen''. He mentions that this data ``could be useful for marketing purposes''. P3 finds \vts potentially useful for helping reproduce hard-to-replicate bugs.

The participants provided valuable and specific feedback for improving \vts. They suggested to enrich the videos produced when executing \vtss script with a bounding box of the GUI components or screen areas being interacted with at each step. They also mention that the video could show (popup) comments that explain what is going on in the app (\eg a comment such as ``after 10 seconds, button X throws an error''), which can help replicate bugs. They would like to see additional information in the script, such as GUI metadata that provides more in-depth and easy-to-read information about each step. For example, the script could use the names or IDs of the GUI components being interacted with and produce steps such as ``the user tapped on the send button'' instead of ``the user tapped at (10.111,34.56)''. P3 mentioned that ``it would be nice to change the script programmatically by using the GUI components' metadata instead of coordinates, so the script is easier to maintain''. They suggest to include an interpreter of commands/steps, written in a simple and easy-to-write language, which would be translated into low-level commands.

\begin{tcolorbox}[enhanced,skin=enhancedmiddle,borderline={1mm}{0mm}{MidnightBlue}]
	Answer to \textbf{RQ$_5$}: Developers find \vts accurate in replicating app usage scenarios from input videos, and potentially useful for supporting several development tasks, including automatically replicating bugs, analyzing usage app behavior, helping users/QA members generate high-quality videos, and automating scenario executions.
\end{tcolorbox}

\section{Limitations \& Threats to Validity}
\label{sec:limitations}
\vspace{0.2cm}
\noindent\textbf{Limitations.} 
Our approach has various limitations that serve as motivation for future work. One current limitation of the \Faster implementation our approach utilizes is that it is tied to the screen size of a single device, and thus a separate model must be trained for each screen size to which \approach is applied. However, as described in \secref{subsec:approach-detection}, the training data process is fully automated and models can be trained once and used for any device with a given screen size. This limitation could be mitigated by increasing dataset size including all type of screen sizes with a trade-off on the training time. To facilitate the use of our model by the research community, we have released our trained models for the two popular screen sizes of the Nexus 5 and Nexus 6P in our online appendix~\cite{appendix}.

Another limitation, which we will address in future work, is that our replayable traces are currently tied to the dimensions of a particular screen, and are not easily human readable. However, combining \approach with existing techniques for device-agnostic test case translation~\cite{Fazzini:ICST17}, and GUI analysis techniques for generating natural language reports~\cite{Moran:ICST16} could mitigate these limitations. 

Finally, as discussed in \secref{subsec:results-rq3}, one limitation that affects the ability of \vts to faithfully replay swipes is the video frame-rate. During the evaluation, our devices were limited to 30fps, which made it difficult to completely resolve a small subset of gesture actions that were performed very quickly. However, this limitation could be addressed by improved Android hardware or software capable of recording video at or above 60fps, which, in our experience, should be enough to resolve nearly all rapid user gestures.

\vspace{0.2cm}
\noindent\textbf{Internal Validity.} In our experiments evaluating \vts, threats to internal validity may arise from our manual validation of the correctness of replayed videos. To mitigate any potential subjectivity or errors, we had at least two authors manually verify the correctness of the replayed scenarios. Furthermore, we have released all of our experimental data and code~\cite{appendix}, to facilitate the reproducibility of our experiments.

\vspace{0.2cm}
\noindent\textbf{Construct Validity.} The main threat to construct validity arises from the potential bias in our manual creation of videos for the popular apps study carried out to answer RQ$_3$. It is possible that the author's knowledge of \vts influenced the manner in which we recorded videos.  To help mitigate this threat, we took care to record videos as naturally as possible (\eg normal speed, included natural quick gestures). Furthermore, we carried out an orthogonal controlled study in the course of answering RQ$_3$, where users unfamiliar with \approach naturally recorded videos on a physical device, representing an unbiased set of videos. 
Another potential confounding factor concerns the quality of the dataset of screens used to train, test, and evaluate \vtss~\Faster and Opacity CNN. To mitigate this threat, we utilize the \ReDraw dataset~\cite{Moran:TSE18} of screens which have undergone several filtering and quality control mechanisms to ensure a diverse set of real GUIs. 
One more potential threat concerns our methodology for assessing the utility of \vts. Our developer interviews only assess the \textit{perceived} usefulness of our technique, determining whether developers actually receive benefit from \vts is left for future work.

\vspace{0.2cm}
\noindent\textbf{External Validity.} Threats to the generalizability of our conclusions are mainly related to: (i) the number and diversity apps used in our evaluation; (ii) the representativeness of usage scenarios depicted in our experimental videos; and (iii) the generalizability of the responses given by the interviewed developers. To help mitigate the first threat, we performed a large-scale study with 64 of the top applications on Google Play mined from 32 categories. While performing additional experiments with more applications is ideal, our experimental set of applications represents a reasonably large number of apps with different functionalities, which illustrate the relatively applicability of \vts. To mitigate the second threat, we collected scenarios illustrating bugs, natural apps usages, real crashes, and controlled crashes from eight participants. Finally, we do not claim that the feedback we received from developers generalizes broadly across industrial teams. However, the positive feedback and suggestions for future work we received in our interviews illustrate the potential practical usefulness of \vts. 

\vspace{-0.2cm}
\section{Related Work}
\label{sec:related-work}
\vspace{0.1cm}

\textbf{Analysis of video and screen captures.}
Lin \etal \cite{Lin:NDSS14} proposed an approach called Screenmilker to automatically extract screenshots of sensitive information (\eg user entering a password) by using the Android Debug Bridge. This technique focuses on the extraction of keyboard inputs from "real-time" screenshots. Screenmilker is primarily focused upon extracting sensitive information, whereas \vts analyzes every single frame of a video to generate a high fidelity replay script from a sequence of video frames.

Krieter \etal \cite{Krieter:MobileHCI18} use video analysis to extract high-level descriptions of events from user video recordings on Android apps. Their approach generates log files that describe what events are happening at the app level. Compared to our work, this technique is not able to produce a script that would automatically replay the actions on a device, but instead simply describe high-level app events (\eg \textit{``WhatsApp chat list closed''}). Moreover, our work focuses on video analysis to help with bug reproduction and generation of test scenarios, rather than describing usage scenarios at a high level.

Bao \etal \cite{Bao:ICSE15} and Frisson \etal \cite{Frisson:CHI16} focus on the extraction of user interactions to facilitate behavioral analysis of developers during programming tasks using CV techniques. In our work, rather than focusing upon recording developers interactions, we instead focus on understanding and extracting generic user actions on mobile apps in order to generate high-fidelity replay scripts.

Other researchers have proposed approaches that focus on the generation of source code for Android applications from screenshots or mock-ups. These approaches rely on techniques that vary solely from CV-based \cite{Nguyen:ASE15} to DL-based \cite{Beltramelli:EICS18,Moran:TSE18,Chen:ICSE18}.

The most related work to \vts is the AppFlow approach introduced by Hu \etal \cite{Hu:FSE18}. AppFlow leverages machine learning techniques to analyze Android screens and categorize types of test cases that could be performed on them (\ie a sign in screen whose test case would be a user attempting to sign in). However, this technique is focused on the generation of semantically meaningful test cases in conjunction with automated dynamic analysis. In contrast, \vts is focused upon the automated replication of any type of user interaction on an Android device, whether this depicts a usage scenario or bug. Thus, \vts could be applied to automatically reproduce crowdsourced mobile app videos, whereas AppFlow is primarily concerned with the generation of tests rather than the reproduction of existing scenarios.

\vspace{0.2cm}
\textbf{Record and replay.}
Many tools assist in recording and replaying tests for mobile platforms \cite{Gomez:ICSE13, airtest, replaykit, Jeon:12,Moran:MOBILESoft'17}. However, many of these tools require the recording of low-level events using \texttt{adb}, which usually requires rooting of a device, or loading a custom operating system (OS) to capture user actions/events that are otherwise not available through standard tools such as \texttt{adb}.
While our approach uses RERAN \cite{Gomez:ICSE13} to replay system-level events, we rely on video frames to transform touch overlays to low-level events. This facilitates bug reporting for users by minimizing the requirement of specialized programs to record and replay user scenarios.

Hu \etal \cite{Hu:OOPSLA15} developed VALERA for replaying device actions, sensor and network inputs (\eg GPS, accelerometer, etc.), event schedules, and inter-app communication. This approach requires a rooted target device and the installation of a modified Android runtime environment. These requirements may lead to practical limitations, such as device configuration overhead and the potential security concerns of rooted devices. Such requirements are often undesirable for developers \cite{Lam:FSE17}. Conversely, our approach is able to work on any unmodified Android version without the necessity of a rooted device, requiring just a screen recording.

Nurmuradov \etal \cite{Nurmuradov:ISSTA17} introduced a record and replay tool for Android applications that captures user interactions by displaying the device screen in a web browser. This technique uses event data captured during the recording process to generate a heatmap that facilitate developers' understanding on how users are interacting with an application. This approach is limited in that users must interact with a virtual Android device through a web application, which could result in unnatural usage patterns. This technique is more focused towards session-based usability testing, whereas \vts is focused upon replaying "in-field" app usages from users or crowdsourced testers collected from real devices via screen recordings.

Other work has focused on capturing high-level interactions in order to replay events \cite{Halpern:ISPASS15, airtest, culebra, espresso}. For instance Mosaic \cite{Halpern:ISPASS15}, uses an intermediate representation of user interactions to make replays device agnostic. Additional tools including HiroMacro \cite{hiromacro} and Barista \cite{Fazzini:ICST17} are Android applications that allow for a user to record and replay interactions. They require the installation or inclusion of underlying frameworks such as \texttt{replaykit} \cite{replaykit}, AirTest \cite{airtest}, or \texttt{troyd} \cite{Jeon:12}. Android Bot Maker \cite{bot_maker} is an Android application that allows for the automation of individual device actions, however, it does not allow for recording high-level interactions, instead one must enter manually the type of action and raw $(x,y)$ coordinates. In contrast to these techniques, one of \vtss primary aims is to create an Android record and replay solution which an inherently low barrier to usage. For instance, there are no frameworks to install, or instrumentation to add, the only input is an easily collectable screen recording. This makes \vts suitable for use in crowd- or beta-testing scenarios, and improves the likelihood of its adoption among developers for automated testing, given its ease of use relative to developer's perceptions of other tools~\cite{Linares-Vasquez:ICSME'17}.

	Finally, as crowdsourcing information from mobile app users has become more common with the advent of a number of testing services~\cite{applause,testbirds,mycrowd}, researchers have turned to utilizing usage data recorded from crowd-testers to enhance techniques related to automated test case generation for mobile apps. Linares-V\'{a}squez~\etal first introduced the notion of recording crowd-sourced data to enhance input generation strategies for mobile testing through the introduction of the {\sc MonkeyLab} framework~\cite{Linares-Vasquez:MSR'15}. This technique adapted N-gram language models trained on recorded user data to predict event sequences for automated mobile test case generation. Mao~\etal developed the {\sc Polariz} approach which is able to infer ``motif'' event sequences collected from the crowd that are applicable across different apps~\cite{Mao:ASE17}. The authors found that the activity-based coverage of the {\sc Sapienz} automated testing tool can be improved through a combination of crowd-based and search-based techniques. The two approaches discussed above require either instrumented or rooted devices ({\sc MonkeyLab}), or interaction with apps via a web-browser ({\sc Polariz}). Thus, \vts is complementary to these techniques as it provides a frictionless mechanism by which developers can collect user data in the form of videos.

\section{Conclusion \& Future Work}
\label{sec:conclusion}

We have presented \approach, an approach for automatically translating video recordings of Android app usages into replayable scenarios.  A comprehensive evaluation indicates that \approach: (i) accurately identifies touch indicators and it is able to differentiate between opacity levels, (ii) is capable of reproducing a high percentage of complete scenarios related to crashes and other bugs, with promising results for general user scenarios as well, and (iii) is potentially useful to support real developers during a variety of tasks. 

Future work can make \approach applicable to different software maintenance tasks, such as: (i) producing scripts with coordinate-agnostic actions, (ii) generating natural language user scenarios, (iii) improving user experience via behavior analysis, (iv) facilitating additional maintenance tasks via GUI-based information, \etc

\begin{acks}

This work is supported in part by the NSF CCF-1927679, CCF-1815186, CNS-1815336, and CCF-1955837 grants. Any opinions, findings, and conclusions expressed herein are the authors' and do not necessarily reflect those of the sponsors.

\end{acks}

\bibliographystyle{ACM-Reference-Format}
\bibliography{minimum}

\end{document}